\documentstyle[prl,aps,multicol]{revtex}
\input epsf
\begin{document}
\draft

\title{Cavity-damping-induced transitions in a driven atom-cavity system}

\author{Hyunchul Nha$^{1,2}$ and Kyungwon An$^2$}
\address{1.School of Physics, Seoul National University, Seoul, Korea\\2.Center for Macroscopic Quantum-Field Lasers and Department of Physics, KAIST, Taejon, Korea}
\maketitle

\begin{abstract}
We investigate the fluorescence spectrum of a two-level atom in a cavity when the atom is driven by a classical field.  We show that forbidden dipole transitions in the Jaynes-Cummings Ladder structure are induced in the presence of the cavity damping, which deteriorates the degree of otherwise perfect destructive interference among the transition channels.  With the larger cavity decay, these transitions are more enhanced.
\end{abstract}


\begin{multicols}{2}

\narrowtext

The radiative properties of the Jaynes-Cummings Molecule (JCM),$^1$ a system composed of a single two-level atom coupled to a single mode of a cavity, need a particular attention.  This system, regarded as one of the simplest and most fundamental forms of matter-radiation interaction,$^2$  has  been extensively studied as a building block for the quantum information processing.$^3$  From the spectrscopic point of view, the dipole transition in the driven JCM is especially interesting because it provides a way to directly measure the eigenenergies of the coupled atom-cavity system. For a resonant atom-cavity system, the JCM has energy eigenvalues given by $n\hbar\omega_0\pm g\sqrt{n}$ with  $n=0,1,2,\ldots$ and $g$ the coupling strength between the atom and the cavity. There have been efforts to experimentally observe such a spectrum of eigen-energies.$^2$ The eigenvalue for $n=1$ with respect to the ground state ($n=0$) has been measured by the experiments on the normal mode 
splitting.$^{4,5}$ Recently, Brune {\it et al}.\ have resolved the time evolution of a two-state Rydberg atom exposed to a microwave radiation in a microwave cavity.$^6$   From the dynamics of the system, the Rabi frequency components corresponding to the eigenvalues for $n\ge1$ were extracted. 

The resonance fluorescence spectrum of JCM  also provides information on the  higher eigenvalues.  Experiments to measure such spectrum in the {\it optical} region are expected in the near future as the  cavity-QED techniques such as the single atom trapping in a cavity are rapidly maturing.$^{7,8}$  In the ideal JCM  where the cavity damping is absent, the dipole transitions only between the adjacent energy manifolds, i.e., between two energy levels with adjacent $n$ values, are possible when the atom is driven by a strong coherent field.   In this work, however, we report that the cavity damping induces new spectral lines corresponding to the transitions that are prohibited in the ideal JCM. One can observe, for example, novel peaks at the frequency $\omega=\omega_0\pm\sqrt{2}g'$,  where $g'$ is the atom-cavity coupling modified by the presence of both the driving field and the cavity damping.  These new peaks correspond to the transition from manifold-2 to manifold-0. The role of the cavity damping in this case is the deterioration of 
the degree of otherwise perfect destructive interference in the transition channels accounting for the transition probability amplitude.  Note that similar transitions can be found in the JCM where the cavity instead of the atom is driven.$^9$ The dynamics imbedded is, however,  quite different from the present one since the anomalous transitions there are allowed by the driving field without the cooperation of the cavity damping. In the present system, the anomalous transitions are more enhanced with the larger cavity damping rate for a given driving-field intensity.  We will derive an effective Hamiltonian which explicitly manifests the dynamic effect of the cavity damping.

In our model a two-level atom coupled with a single-mode cavity is driven by a resonant classical field, as described by an interaction Hamiltonian 
\begin{eqnarray}
H_I=i\hbar g (a^{\dag}\sigma_{-}-a\sigma_{+})+H_d
\label{eq1}
\end{eqnarray}
where
\begin{eqnarray}
H_d&=&i\hbar\Omega(\sigma_+-\sigma_-).
\label{eq2}
\end{eqnarray}
The atom-cavity coupling strength is denoted by $g$ while $\Omega$ is the Rabi frequency of the driving field, $\sigma_{\pm}$ and $\sigma_z$ the atomic pseudospin operators and ${a}^{\dag}(a)$ the creation (annihilation) operator of the cavity mode. For comparison, when the cavity instead of the atom is driven, $H_d$ is replaced with the form
\begin{eqnarray}
H_d&=&i\hbar{\cal E}(a^{\dag}-a),
\label{eq3}
\end{eqnarray}
which will be referred later.

Let us first consider the undamped system. The eigenenergies and the eigenstates of $H_I$ have been obtained by Alsing {\it et al.}$^{10}$ as 
\begin{eqnarray}
E_{n}^{\pm}=\pm\hbar\sqrt{n}g,(n=0,1,2,...)
\label{eq4}
\end{eqnarray}
and
\begin{eqnarray}
&|&\phi_0\rangle=D(\alpha)|0\rangle|-\rangle,\nonumber\\
&|&\phi_n^{\pm}\rangle=D(\alpha)\frac{1}{\sqrt{2}}\left[|n-1\rangle|+\rangle\pm i|n\rangle|-\rangle\right]
\label{eq5}
\end{eqnarray}
where $\alpha\equiv\Omega/g$, and $D(\alpha)$ denotes the displacement operator for the cavity-field mode. The notation $|n_c\rangle|\pm\rangle$ represents the state in which the cavity photon number is $n_c$ and the atom is in its excited/ground state. 

In the cavity-driving case, however, the quasi-energies depend on the driving field as$^{10}$ 
\begin{eqnarray}
E_{n}^{\pm}=\pm\hbar\sqrt{n}g[1-(2{\cal E}/g)^2]^{3/4}\equiv\pm\hbar\sqrt{n}g',
\label{eq6}
\end{eqnarray}
where $n=0,1,2,\cdots.$
We have found in the previous work$^9$ that the anomalous dipole-transition lines are found in the spectrum of the resonance fluorescence at $\omega=\omega_0\pm\sqrt{2}g',\pm\sqrt{3}g'$, etc. due to the dynamic effect induced by the driving field in the cavity-driving case. In that case, the positions of the transition lines in the spectrum are highly dependent on the driving field intensity since $g'=g[1-(2{\cal E}/g)^2]^{3/4}$.  In the atom-driving case, by a similar argument, one might expect similar transition lines might be induced by the strong driving field, at $\omega=\omega_0\pm\sqrt{2}g,\pm\sqrt{3}g$ etc..  We show below, however, that the driving-field alone cannot induce such transitions.  

In the absence of atom/cavity damping, the matrix elements for the dipole transitions from the quasi-energy state $|\phi_n^{\xi}\rangle$ to $|\phi_{n'}^{\xi'}\rangle$ are calculated as 
\begin{eqnarray}
\langle\phi_{n'}^{\xi'}|\sigma_-|\phi_n^{\xi}\rangle
=i^{\xi'}\delta_{n',n-1} \;,
\label{eq7}
\end{eqnarray}
where
$\xi,\xi'=\pm1$ stand for the upper(+) and the lower(-) states, respectively.
From the above selection rule, it is clear that only the transitions between adjacent manifolds (from $n$ to $n-1$) are possible.

Now, let us turn our attention to the realistic situation where both the atomic and the cavity damping are included. The time evolution of the system is obtained by the master equation for the density operator $\rho_I$ as
\begin{eqnarray}
\dot{\rho_I}&=&[g(a^{+}\sigma_{-}-a\sigma_{+})+\Omega(\sigma_+-\sigma_-),\rho_I]\nonumber\\
&+&(\gamma/2)(2\sigma_{-}\rho_I\sigma_{+}-\sigma_{+}\sigma_{-}\rho_I-\rho_I\sigma_{+}\sigma_{-})\nonumber\\
&+&\kappa(2a\rho_I a^{+}-a^{+}a\rho_I-\rho_I a^{+}a)
\label{eq9}
\end{eqnarray}
where $\gamma$ the decay rate of the atom into free-space vacuum-field modes and $2\kappa$ the cavity damping rate.  If the atom is inside an ideal lossless cavity ($\kappa=0$), in the steady state the cavity field becomes a coherent state and the atom remains its ground state, i.e., $\rho_I^{ss}=|\alpha,-\rangle\langle\alpha,-|$. Thus, in this case the fluorescence of the atom is suppressed regardless of the strength of $\gamma$.$^{11}$ The cavity in real experiments, however, are always lossy, and thus the atom can be excited and the fluorescence is not suppressed.  

We have numerically calculated the fluorescence spectrum using the quantum regression theorem$^{12}$including the cavity damping. The results are shown  in Fig.\,1, where the dominant peak {\it a} is found near at $\omega_0+g$ (and its counterpart at  $\omega_0-g$), which corresponds to the usual vacuum Rabi-splitting.  Interestingly, additional peaks {\it b} and {\it c} are found near at $\omega_0+\sqrt{2}g$ and $\omega_0+2g$, respectively. As discussed before based on Eq.\,(\ref{eq7}), they are the forbidden transitions when $\kappa=0$. These anomalous peaks are enhanced as the driving-field intensity is increased further as in Fig.\,1 (b).  The peak {\it d} near at $(\sqrt{2}-1)g$, on the other hand, is the allowed one.  
The novel dynamics imbedded in these anomalous transitions is induced by the presence of the cavity damping in addition to the driving field.

The master equation, Eq.\,(\ref{eq9}) is not much instructive in the present form to understand the dynamics, so it is transformed using $\tilde{\rho}=D^+(\alpha)\rho_ID(\alpha)$ into
\begin{eqnarray}
\dot{\tilde{\rho}}&=&[g(a^{+}\sigma_{-}-a\sigma_{+}),\tilde{\rho}]+\Omega\frac{\kappa}{g}[a^+-a,\tilde{\rho}]\nonumber\\
&+&(\gamma/2)(2\sigma_{-}\tilde{\rho}\sigma_{+}-\sigma_{+}\sigma_{-}\tilde{\rho}-\tilde{\rho}\sigma_{+}\sigma_{-})\nonumber\\
&+&\kappa(2a\tilde{\rho} a^{+}-a^{+}a\tilde{\rho}-\tilde{\rho} a^{+}a)
\label{eq10}
\end{eqnarray}
In this `displaced' frame, 
the effective Hamiltonian is derived from the above expression as
\begin{eqnarray}
H_{\rm eff}=i\hbar g (a^{+}\sigma_{-}-a\sigma_{+})+i\hbar\Omega\frac{\kappa}{g}(a^+-a)
\label{eq11}
\end{eqnarray} 
This is the very interaction Hamiltonian describing the cavity-driving case (see  \,Eq.\,(\ref{eq3})) with the effective driving field ${\cal E}_{\rm eff}=\Omega\kappa/g$.  Once we make this identification,  all the peculiar features in the fluorescence spectra can be explained. The eigenvalues of $H_{\rm eff}$ for nonzero $\kappa$ are given by replacing ${\cal E}$ with ${\cal E}_{\rm eff}=\Omega\kappa/g$ in Eq.\,(\ref{eq6}) as
\begin{eqnarray}
E_{n}^{\pm}=\pm\hbar\sqrt{n}g[1-(2\Omega\kappa/g^2)^2]^{3/4},(n=0,1,2,...),
\label{eq12}
\end{eqnarray}
and the corresponding eigenstates $|\chi_n^{\pm}\rangle$ are similarly obtained. As shown in Ref.\,9, the transitions between the non-adjacent manifolds are possible in the cavity-driven JCM. For example, the peak {\it b} corresponds to the transition from $|\chi_2^+\rangle$ to $|\chi_0\rangle$. To the first order in ${\cal E}_{\rm eff}/g$, $|\chi_0\rangle \approx$ $|0,-\rangle-({\cal E}_{\rm eff}/g) |0,+\rangle$, and $|\chi_2^+\rangle\approx|1,+\rangle-2\sqrt{2}({\cal E}_{\rm eff}/g)[|0,+\rangle+\sqrt{2}|2,+\rangle]\cdots$. Thus, the transition channel from $|0,+\rangle$ to $|0,-\rangle$ gives rise to the transition matrix element to the lowest order in $2\Omega\kappa/g^2$ as
\begin{eqnarray}
|\langle\chi_0|\sigma_-|\chi_2^{+}\rangle|^2=(2\Omega\kappa/g^2)^2.
\label{eq13}
\end{eqnarray} 
Note that in case of the lossless cavity with $\kappa=0$, the strength of the transition is zero. Thus, we
conclude that the anomalous transitions are induced by the cavity decay in cooperation with the real driving field.  In addition, from Eq.\,(\ref{eq12}),  it can be easily deduced that the effective atom-cavity coupling constant is not $g$ but $g'\equiv g[1-(2\Omega\kappa/g^2)^2]^{3/4}$, so the exact positions of the peaks {\it a, b, c} are predicted to be at $g', \sqrt{2}g', 2g'$, respectively, which is numerically confirmed in Fig.\,1.

We can gain further physical insight on the dynamic effect of the cavity damping by expanding the transition matrix elements in the increasing orders of $\alpha$.  For instance, in the absence of the damping, the transition matrix element for the peak at $\sqrt{2}g'$ is
given as
\begin{eqnarray}
\langle\phi_0|\sigma_-|\phi_2^+\rangle&=&\alpha\left[-\frac{i}{2}\langle0,-|\sigma_-|0,+\rangle+\frac{i}{2}\langle1,-|\sigma_-|1,+\rangle\right]\nonumber\\&+&O(\alpha^2)+\cdots=0.
\label{eq14}
\end{eqnarray}
\begin{figure}
\centerline{\epsfxsize=8cm\epsffile{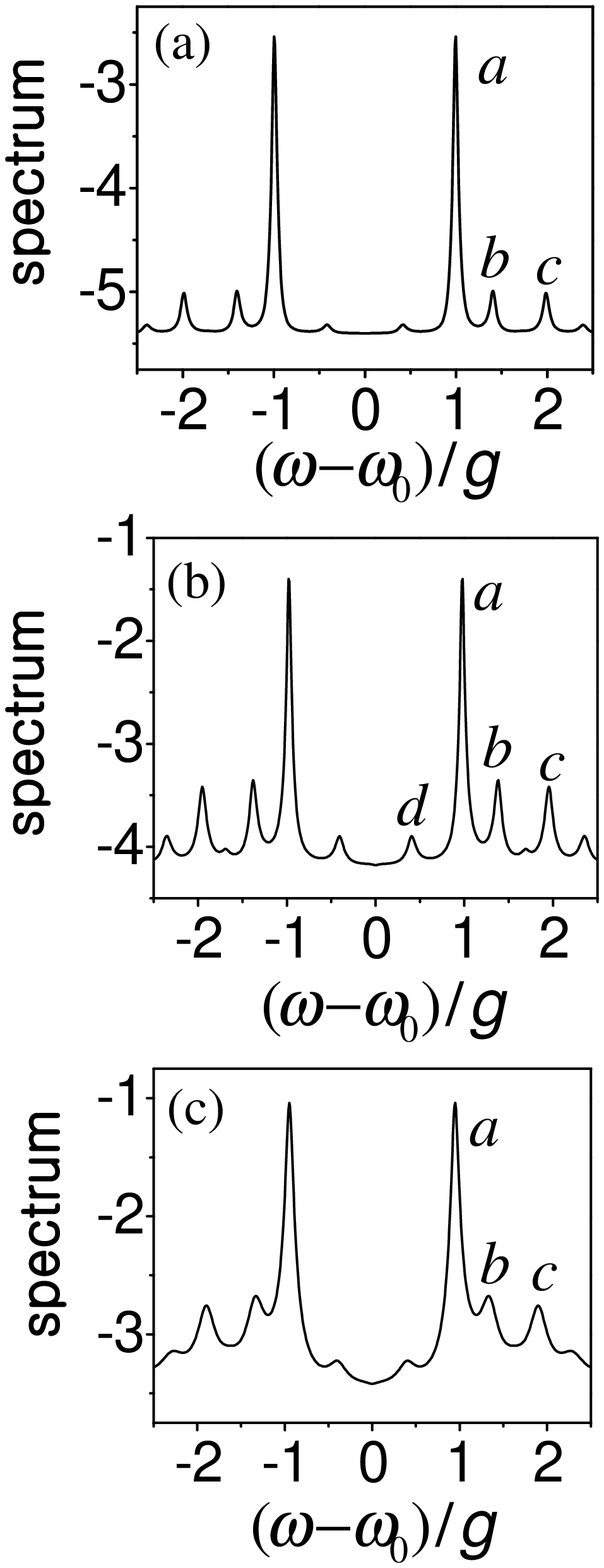}}
\caption{The spectrum (in arbitrary unit and in log scale) of the atomic fluorescence.  (a) $\Omega/g=8/3$, $2\kappa/g=0.03$, (b) $\Omega/g=16/3$,  $2\kappa/g=0.03$,  and (c) $\Omega/g=8/3$, $2\kappa/g=0.1$. Location of each peak is identified as ${\it a}\to g'$, ${\it b}\to \sqrt{2}g'$, ${\it c}\to 2g'$, ${\it d}\to (\sqrt{2}-1)g'$ with $g'=g[1-(2\Omega\kappa/g^2)^2]^{3/4}$.}
\end{figure}

Although the driving field induces two transition channels $\langle0,-|\sigma_-|0,+\rangle$ and $\langle1,-|\sigma_-|1,+\rangle$ to the first order of $\alpha$, the probability amplitudes associated with these channels have the same magnitude and the opposite sign, so that they are exactly cancelled. Due to this destructive interference, the transition is forbidden.  Similar arguments can be made for all orders of $\alpha$.  When the cavity damping is introduced, however, the cancellation is not exact, and thus the amplitude does not vanish.  In this case, each number state $|n\rangle$ $(n=1,2,\cdots)$,  except the vacuum state $|0\rangle$, appearing in the above transition matrix element, undergoes `broadening',  containing other number-state components, so that two channels do not cancel each other exactly.  As the driving field gets stronger the higher order terms in Eq. (\ref{eq14}) also become significant.  Furthermore, as the cavity damping deteriorates the destructive interference between the channels, the larger the cavity damping rate, the stronger the deterioration. Therefore, it is now explained why the effective driving field ${\cal E}_{\rm eff}$, which causes the anomalous transitions, is proportional to both the driving field intensity and the cavity damping rate. Similarly, the other anomalous peaks can also be explained.
  
The dynamic effect of the cavity-damping can be seen more dramatically when the cavity-damping rate is increased while the driving-field Rabi frequency is fixed.  As confirmed in Fig.\,1 (c),  the anomalous transitions are enhanced more compared with those in Fig.\,1(a).  Of course, there is trade-off in this way of enhancing anomalous peaks because of the usual line broadening effect of cavity damping.  In addition, the cavity damping cannot be increased beyond $\kappa=g^2 / 2\Omega$.  Otherwise, the quasi-energy-state  description is no longer valid as Eq.\,(\ref{eq12}) implies. 

In conclusion, we have investigated the anomalous spectral lines in the fluorescence spectrum of JCM in the atom-driving case.  We have shown that the forbidden transitions with the driving field alone can be allowed by the cavity damping and that these transitions are enhanced as the damping rate is increased.  This novel effect of cavity damping in the dynamics of JCM can also be observed through similar transition lines in the spectrum of the cavity transmission although only the results of atomic fluorescence are presented here.
 
This work is supported by Creative Research Initiatives of the Korean Ministry of Science and Technology.
\vspace{-0.8cm}
\bibliographystyle{prsty}

\end{multicols} 
\end{document}